\documentclass[aps,twocolumn,showpacs,showkeys,amsmath,amssymb,nofootinbib,superscriptaddress,prc,floatfix]{revtex4-1}
\usepackage{graphicx}
\usepackage{subfigure}
\usepackage{bm}

\makeatletter
\newcommand\erfc{\mathop{\operator@font erfc}\nolimits}
\def\slashchar#1{\setbox0=\hbox{$#1$}
   \dimen0=\wd0 \setbox1=\hbox{/} \dimen1=\wd1
   \ifdim\dimen0>\dimen1 \rlap{\hbox to \dimen0{\hfil/\hfil}} #1
   \else  \rlap{\hbox to \dimen1{\hfil$#1$\hfil}} / \fi}

\begin{document}
\title{Particle spectra in Pb-Pb collisions at $\sqrt{s_{NN}}=2.76$~TeV}
\author{Piotr Bo\.zek}
\affiliation{
Institute of Physics, Rzesz\'ow University, 
PL-35959 Rzesz\'ow, Poland}
\affiliation{The H. Niewodnicza\'nski Institute of Nuclear Physics,
PL-31342 Krak\'ow, Poland} 
\author{Iwona Wyskiel-Piekarska}
\affiliation{The H. Niewodnicza\'nski Institute of Nuclear Physics,
PL-31342 Krak\'ow, Poland}
\date{\today}

\begin{abstract}
Particle production  in Pb-Pb collisions at $\sqrt{s_{NN}}=2.76$TeV is studied in the $(3+1)$-dimensional 
viscous hydrodynamic model.  The shapes of the calculated 
transverse momentum spectra of $\pi^{+}$, $K^{+}$, protons, $\Xi^{-}$, 
and
$\Omega^{-}$ are in satisfactory agreement with preliminary data of the ALICE Collaboration, while the particle ratio $p/\pi^+$ is slightly overpredicted, 
and the strange barion yields are underpredicted.
\end{abstract}

\pacs{25.75.Ld, 24.10Nz, 24.10Pa}

\keywords{relativistic 
heavy-ion collisions,  hydrodynamic model, transverse momentum spectra}

\maketitle

\section{Introduction}

Particle production in Pb-Pb collisions at the highest available energy $\sqrt{s_{NN}}=2.76$~TeV has been
studied experimentally at the CERN Large Hadron Collider (LHC) \cite{Aamodt:2010cz,*Toia:2011rt,*ATLAS:2011ah,*Aad:2010bu,*Aad:2012bu,*Chatrchyan:2011pb,*Chatrchyan:2011sx,*Aamodt:2011vk,Aamodt:2011mr,Floris:2011ru}. 
The observation of the elliptic and triangular flows  indicates that a collectively expanding fireball
of dense matter is formed, confirming results obtained in collisions at lower energies 
\cite{Arsene:2004fa,*Back:2004je,*Adams:2005dq,*Adcox:2004mh}.  The hydrodynamic model 
of the  dynamics  provides a quantitative explanation for  observables defined for
particles emitted with soft momenta
 \cite{Kolb:2003dz,*Huovinen:2006jp,*Florkowski:2010zz}. 
In particular, hydrodynamic models are applied to describe the  anisotropic flow 
 of charged particles produced in Pb-Pb collisions at the LHC 
\cite{Luzum:2010ag,Hirano:2010je,Shen:2011eg,Song:2011qa,Bozek:2011wa,Schenke:2011tv,Qiu:2011hf,Retinskaya:2012ky,Niemi:2012ry}.

Statistical models of the particle production in  heavy-ion collisions predict the production 
rates  of specific 
hadrons  assuming a chemically equilibrated system 
\cite{BraunMunzinger:2001ip,*Andronic:2005yp,*Cleymans:2004pp,*Florkowski:2001fp,*Rafelski:2004dp,*Becattini:2005xt}. Recent results for  Pb-Pb collisions at the LHC seem to be incompatible with this simple mechanism.
The number of protons emitted is lower than expected from the  rates of the emission of 
other particles, which raises a doubt  in the assumption that the production of 
different particle species  happens at 
a common chemical freeze-out temperature \cite{Preghenella:2011jv}. 
Particle abundances may undergo significant modifications in the nonequilibrium dynamics
after hadronization, e.g.  annihilation processes may reduce baryon  multiplicities
\cite{Becattini:2012sq,*Steinheimer:2012rd}. 

Transverse momentum spectra of identified particles represent a more
basic observable, as they involve the overall multiplicity as well as the momentum distribution for
each particle species. The hydrodynamic evolution of a fireball of 
 thermally equilibrated fluid 
until the freeze-out temperature $T_f$, leads to a common chemical and kinetic freeze-out temperature. 
A similar idea constitutes the basic assumption of the single freeze-out model of particle emission 
\cite{Broniowski:2001we,*Rybczynski:2012ed}. Additional freedom is allowed in hydrodynamic models assuming
that below the chemical freeze-out temperature the matter is in kinetic equilibrium but
particle abundances remain frozen \cite{Hirano:2002ds}. In such models particle ratios 
correspond to a fixed chemical freeze-out temperature $T_{chem}$, while the transverse momentum spectra 
are determined by the convolution of the collective velocity with the thermal emission at the kinetic 
freeze-out temperature $T_{kin}$.  The genuinely nonequilibrium phase at the end of the evolution may
 be addressed using a hybrid model with hydrodynamics
  for the dense phase and  a hadronic cascade afterburner 
for the latter evolution \cite{Bass:2000ib,Hirano:2007xd,Werner:2009fa,Song:2010aq,Petersen:2011sb}. 
The evolution in the hadronic cascade  changes the chemical composition in the system 
and the momentum spectra of  particles.

Deviations from local equilibrium in the hydrodynamic model of the dynamics are introduced as viscosity 
corrections to the energy-momentum tensor
\cite{IS,Teaney:2003kp,Song:2007ux,Dusling:2009df,Chaudhuri:2006jd,Dusling:2007gi,Romatschke:2009im,Teaney:2009qa,Luzum:2008cw,Bozek:2009dw,Schenke:2010rr}. In particular, shear viscosity is important 
in quantitative predictions for the elliptic and triangular collective flows. 
The rapid expansion of the fireball introduces sizable corrections from bulk viscosity, if the 
equilibration processes are not fast enough to restore local equilibrium. Such deviation are twofold.
First, the chemical composition remains effectively frozen at some stage, while the local energy density
drops. Second, the local momentum distributions of particles in the fluid become softer. 
Bulk viscosity leads to both effects, depending on the 
 local expansion rate   \cite{Monnai:2009ad,Bozek:2009dw,Bozek:2011ua,Dusling:2011fd}. 

We present calculations for the transverse momentum spectra of identified particles in a $(3+1)$-dimensional
$[(3+1)$-D$]$ hydrodynamic model with bulk and shear viscosities for Pb-Pb collisions at LHC energies.
 The presence of 
bulk viscosity in the hadronic phase yields nonequilibrium effects for the  chemical composition 
and for the transverse momentum spectra. We find a satisfactory agreement with preliminary data 
of the ALICE Collaboration \cite{Floris:2011ru,Preghenella:2011jv} for pion, kaon, proton  spectra and abundances.
At the same time we constrain the freeze-out temperature using the interferometry data 
\cite{Aamodt:2011mr}. Reproducing the charged particle density in pseudorapidity \cite{Floris:2011ru} 
and the transverse momentum spectra allows for a direct prediction on the rapidity distributions 
of identified hadrons.

\section{Initial conditions and hydrodynamic evolution}

The expansion of the fireball is described using second order viscous hydrodynamics.
Hydrodynamic equations are solved in $(3+1)$-D together with the 
Israel-Stewart equations for
the stress corrections $\pi^{\mu\nu}$ and $\Pi$ to the energy-momentum tensor (for details see \cite{Bozek:2011ua}). 
We use a constant ratio of shear viscosity to 
entropy density  $\eta/s=0.08$. The bulk viscosity is nonzero only in the hadronic phase,
 we use $\zeta/s=0.04$ and $\zeta/s=0.08$. The equation of state is a combination of the lattice
QCD  \cite{Borsanyi:2010cj} and  hadron gas equations of state, obtained in a thermodynamically 
consistent way \cite{Chojnacki:2007jc}. The initial time for the hydrodynamic expansion is $0.6$~fm/c.
The  relaxation times in the Israel-Stewart equations are  $\tau_\pi=\tau_\Pi=\frac{3\eta}{T s}$. For the chosen values
of the relaxation times the stress corrections are close to the ones from the 
Navier-Stokes expression, at latter stages
of the expansion. For a strong temperature dependence of the viscosity coefficients, the value of the relaxations time 
can influence the evolution \cite{Song:2009rh}.

The initial entropy density $s(\eta_\parallel,x,y)$ 
 for the $(3+1)$-D hydrodynamic
 evolution
in the space-time rapidity $\eta_\parallel$ 
and the transverse coordinates $x,\ y$ is
\begin{eqnarray}
s(\eta_\parallel,x,y)  & \propto& \left( \frac{(y_b+\eta_\parallel)N_+
+(y_b-\eta_\parallel)N_-}{y_b (N_++N_-)}\right)\nonumber \\
 & & \left[ \frac{1-\alpha}{2}\rho_{part} +\alpha \rho_{bin}\right] f(\eta_\parallel) .
\label{eq:ini}
\end{eqnarray}
In the transverse plane the density is defined as a combination 
of the 
participant nucleon  $\rho_{part}=N_++N_-$ and binary collision 
$\rho_{bin}$ densities. 
The factor $\left( \frac{(y_b+\eta_\parallel)N_+
+(y_b-\eta_\parallel)N_-}{y_b (N_++N_-)}\right)$ in Eq. \ref{eq:ini} implements in  the initial density 
the  assumption that forward (backward) going nucleons $N_+$ ($N_-$) emit particles preferentially 
in the forward (backward) rapidity hemisphere \cite{Bialas:2004su}.
The parameters of  the longitudinal profile 
\begin{equation}
f(\eta_\parallel)\exp\left(-\frac{(\eta_\parallel-\eta_0)^2}{2\sigma_\eta^2}\theta(|\eta_\parallel|-\eta_0)
\right)
\end{equation}
are adjusted to reproduce the  charge particle density in pseudorapidity, $\eta_0=2.3$ and 
$\sigma_\eta=1.4$, $y_b$ is the beam rapidity.
The parameters of the Glauber model used to calculate the entropy profiles  for Pb-Pb collisions 
($A=208$)  are
  $R_A=6.48$fm and
$a=0.535$fm; the nucleon-nucleon cross section is $\sigma=62$mb.

At the freeze-out temperature particles are emitted from the freeze-out hypersurface
according to the Cooper-Frye formula with viscosity corrections. The nonequilibrium 
modifications of the equilibrium momentum distribution $f_0$ 
from shear viscosity are quadratic in momentum
 \cite{Teaney:2003kp}
\begin{equation}
\delta f_{shear}= f_0
\left(1\pm f_0 \right) \frac{1}{2 T^2 (\epsilon+p)}p^\mu p^\nu \pi_{\mu\nu} .
\label{eq:dfsh}
\end{equation}
The correction from bulk viscosity are taken from the relaxation time formula
\cite{Gavin:1985ph,*Hosoya:1983xm,*Sasaki:2008fg,Bozek:2009dw}
\begin{equation}
\delta f_{bulk}= C_{bulk}f_0
\left(1\pm f_0 \right)\left(c_s^2 u^\mu p_\mu -\frac{(u^\mu p_\mu)^2-m^2}
{3 u^\mu p_\mu}\right) \Pi   ,
\label{eq:bcorr}
\end{equation}
where $c_s$ is the sound velocity and $C_{bulk}$ is a  normalization constant. At 
the freeze-out hypersurface,
the flow velocity and the stress corrections from viscosity are exported to  a 
Monte-Carlo statistical emission code \cite{Chojnacki:2011hb} that is used to generate particle spectra.
Bulk viscosity corrections are calculated with respect to the equilibrium distribution having the same energy density. 
Imposing constraints corresponding to  conserved charges leads to deviations from the chemical equilibrium with respect to
this equilibrium reference state. Effectively, it means that due to incomplete equilibration the chemical equilibrium temperature is shifted up 
and  the effective kinetic temperature is shifted down  (due to a redshifting of particle  thermal momenta). An expanded discussion 
of this approach can be found in \cite{Bozek:2011ua,Dusling:2011fd}.

\section{Results}

\begin{figure}
\includegraphics[width=.44\textwidth]{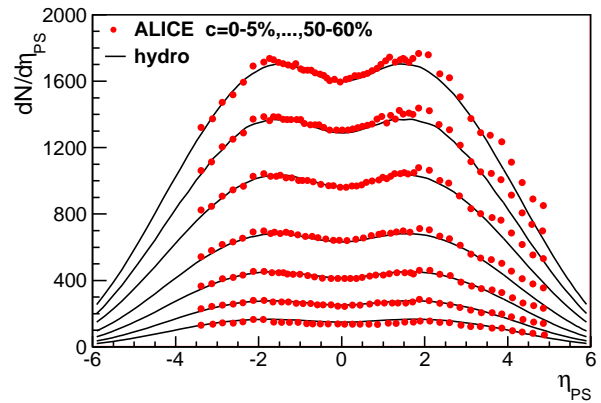}
\caption{(Color online) Charge particle distributions in pseudrapidity in Pb-Pb collision at
$\sqrt{s}_{NN}=2.76$~TeV, for centralities (from top to bottom) $0-5$\%, $5-10$\%, $10-20$\%, 
\dots, $50-60$\%.  Results from viscous hydrodynamics (lines) are compared to preliminary data of the ALICE
Collaboration (symbols) \cite{Floris:2011ru}.}
\label{fig:dndeta}
\end{figure}

\begin{figure}
\includegraphics[width=.44\textwidth]{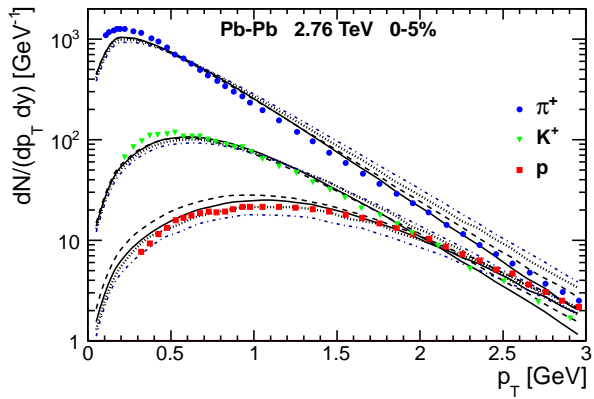}
\caption{(Color online) Transverse momentum spectra of  $\pi^+$, $K^+$, p (from top to bottom) in Pb-Pb collisions with centralities $0-5$\%, obtained
  in the  viscous hydrodynamic model with $\zeta/s=0.04$ and $T_f=150$~MeV (dashed lines), $\zeta/s=0.04$ and $T_f=140$~MeV (dotted lines), $\zeta/s=0.08$ and $T_f=140$~MeV (solid lines). Symbols represent preliminary data of the ALICE Collaboration \cite{Floris:2011ru}. The dashed-doted line represents the result of the viscous hydrodynamic calculation with  $\zeta/s=0.08$ and $T_f=140$~MeV but without bulk viscosity corrections at freeze-out. }
\label{fig:p05}
\end{figure}

The parameters of the initial entropy density for the hydrodynamic evolution are constrained 
by the charge particle density as function of pseudorapidity measured for different centrality classes 
(Fig. \ref{fig:dndeta}). 
We find  that the optimal  mixing parameter for the admixture of binary collisions $\alpha=0.15$;
the same value as used in $(2+1)$-D viscous model simulations for  heavy-ion collisions at the LHC
 \cite{Bozek:2011wa}. The half-width of the plateau in the distribution $\eta_0=2.3$ is larger 
as compared to $\eta_0=1.5$ that has been  used for Au-Au collisions at $\sqrt{s_{NN}}=200$~GeV
 \cite{Bozek:2011ua}.

\begin{figure}
\includegraphics[width=.42\textwidth]{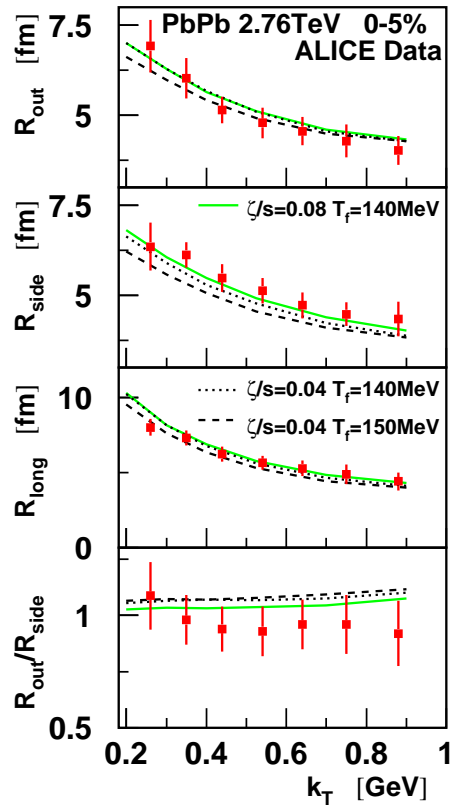}
\caption{(Color online) Interferometry  radii in Pb-Pb collisions with centralities $0-5$\%  
as functions of the pion pair 
momentum,  obtained
  in the  viscous hydrodynamic model with $\zeta/s=0.04$ and $T_f=150$~MeV (dashed lines), $\zeta/s=0.04$ and $T_f=140$~MeV (dotted line), $\zeta/s=0.08$ and $T_f=140$~MeV (solid lines). 
Symbols represent preliminary data of the ALICE Collaboration \cite{Aamodt:2011mr}. }
\label{fig:hbt}
\end{figure}

Transverse momentum spectra for $p^+$, $K^+$ and protons in central collisions are 
presented in Fig. \ref{fig:p05}. The spectra get harder when  the freeze-out temperature is lowered
 from $T_f=150$~MeV (dashed lines) to  $140$~MeV (dotted lines), and using $\zeta/s=0.04$. 
Especially, the pion spectra get too flat. The size of the fireball at freeze-out can be estimated from
the interferometry radii \cite{Heinz:1999rw,*Wiedemann:1999qn,*Weiner:1999th,*Lisa:2005dd}.
The radius $R_{side}$ is best described using a freeze-out temperature of 
$140$~MeV (Fig. \ref{fig:hbt}).
Also the value of the ratio $R_{out}/R_{side}$ is closer to the 
experimental data if the  evolution is longer.
The proton yield is too large for the freeze-out at $150$~MeV, while the average transverse 
momentum of protons is too low. The proton spectra and yields are better reproduced for $T_f=140$~MeV. 
To describe at the same time the pion and proton spectra a freeze-out temperature $140$~MeV 
and a bulk viscosity coefficient $\zeta/s=0.08$  are used 
(solid lines in Figs. \ref{fig:p05} and \ref{fig:hbt}). Increasing the bulk viscosity makes
the local momentum distributions of light particles (pions) softer, which results in the softening of 
their 
final transverse momentum spectra. At the same time, non-equilibrium corrections  make the proton 
yield to increase slightly  as compared to the case of $\zeta/s=0.04$, without changing the proton average
 momentum. In the following we set $T_f=140$~MeV and $\zeta/s=0.08$ as the optimal parameters for the 
freeze-out conditions that reproduce particle spectra and interferometry radii in central collisions.
 We note, that the number of pions and kaons with very soft momenta is underestimated, while the proton yield is slightly overestimated.

\begin{figure}
\includegraphics[width=.44\textwidth]{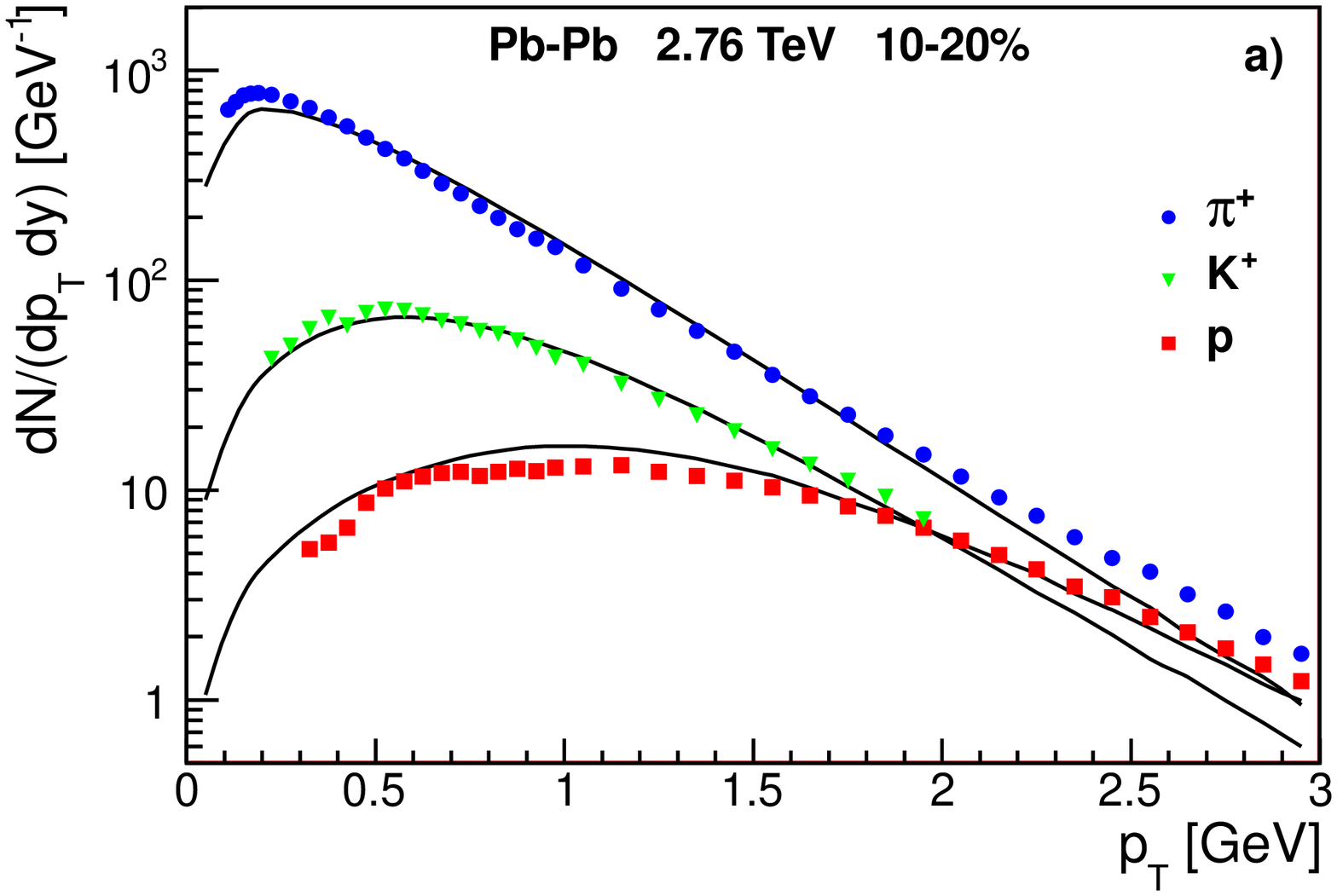}
\includegraphics[width=.44\textwidth]{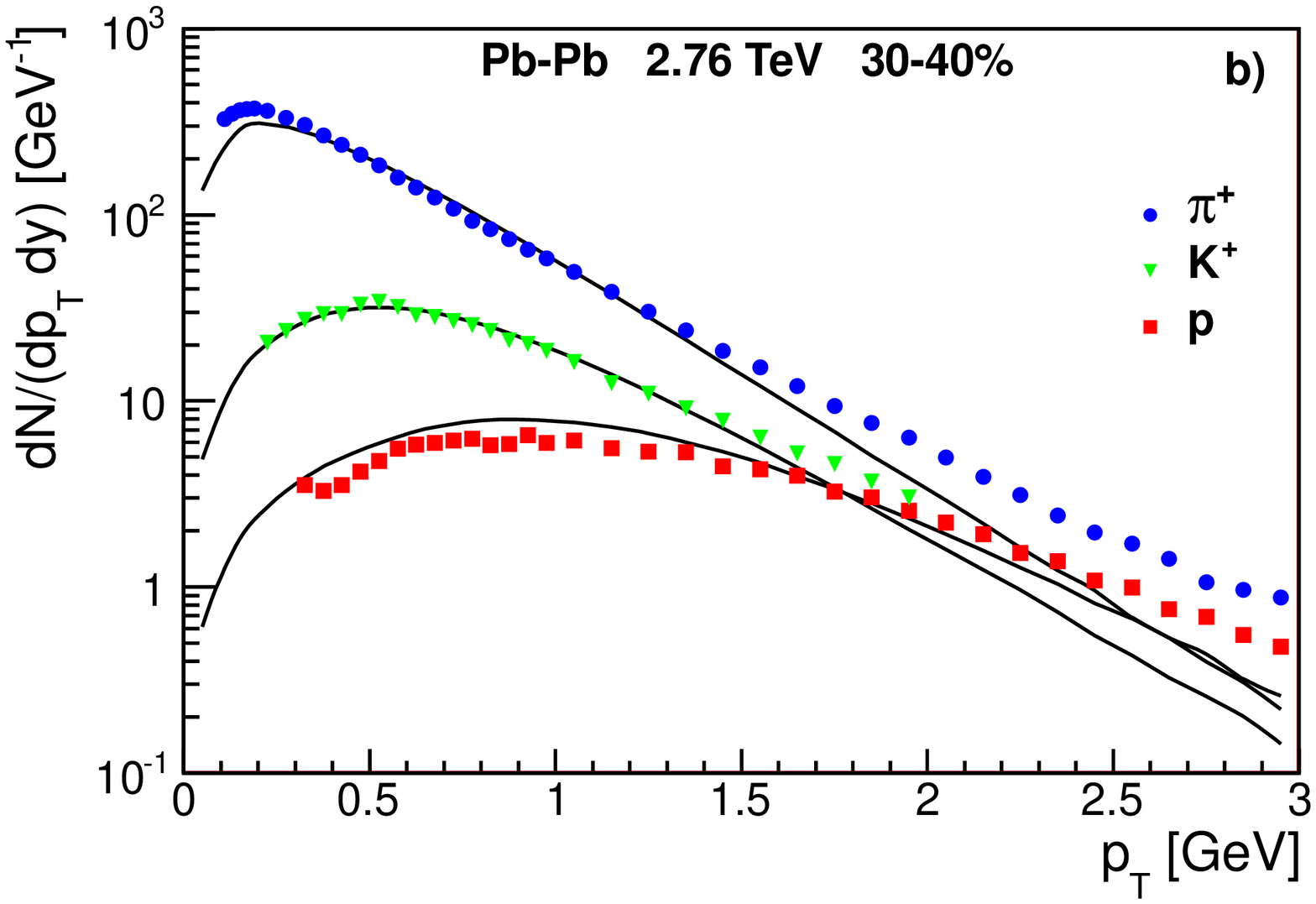}
\caption{(Color online) Transverse momentum spectra of  $\pi^+$, $K^+$, p (from to to bottom)  in
 Pb-Pb collisions with centralities $10-20$\% (panel a) and  $30-40$\% (panel b), 
obtained in the  viscous hydrodynamic model (solid lines). 
Symbols represent preliminary data of the ALICE Collaboration \cite{Floris:2011ru}. }
\label{fig:p1040}
\end{figure}

The pion, kaon, and proton spectra without the bulk viscosity corrections at freeze-out 
 (Eq. \ref{eq:bcorr}) are shown by the dashed-dotted lines in Fig. \ref{fig:p05}. Bulk viscosity corrections make the 
spectra softer, especially for light particles, and introduce a correction in particle abundances, increasing the proton number.
The corrections are
 important for pions with high momenta ($p_\perp>1.5$GeV). They make  the proton number to increase 
by $35$\%. If the
 corrections are large, formally a better ansatz for the distribution function with 
 bulk viscosity corrections would be an exponential function
\cite{Bozek:2011ua}, but the final spectra are very similar as when using Eq. \ref{eq:bcorr}.
The shift in the effective chemical equilibrium  temperature is approximately  the same for all the particles, depending only on the 
local expansion rate. A more elaborate ansatz is possible, with different bulk viscosity corrections for mesons, barions, 
or strange particles \cite{Dusling:2011fd}, using additional parameters the measured particle ratios could be better reproduced.

The spectra of identified particles in semi central collisions are well described by the hydrodynamic model
(Fig. \ref{fig:p1040}).  For momenta 
$p_T>1.5$~GeV the pion spectra are underestimated by the hydrodynamic model with statistical particle 
emission. The discrepancy increases with centrality, and is visible for kaons as well for 
centralities $30-40$\%. This effect indicates that a nonthermal component in the particle emission is 
present, e.g.  jet fragmentation. A similar underestimation of the experimental particle 
yields at high momenta by the hydrodynamic model is seen in peripheral 
Au-Au collisions at lower energies \cite{Bozek:2011ua}. The proton multiplicity and spectra 
are well described by the model for different centrality classes.

\begin{figure}
\includegraphics[width=.44\textwidth]{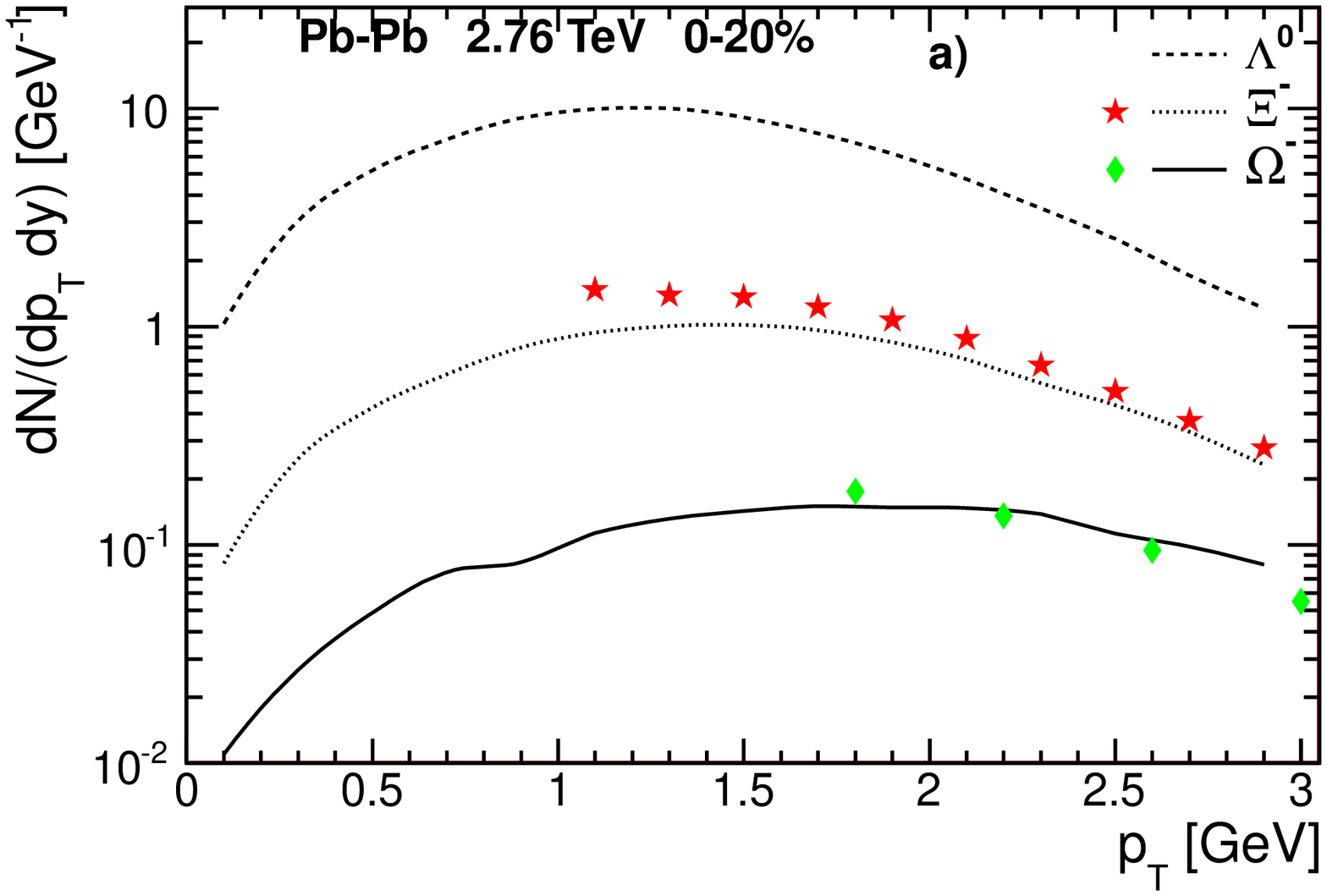}
\includegraphics[width=.44\textwidth]{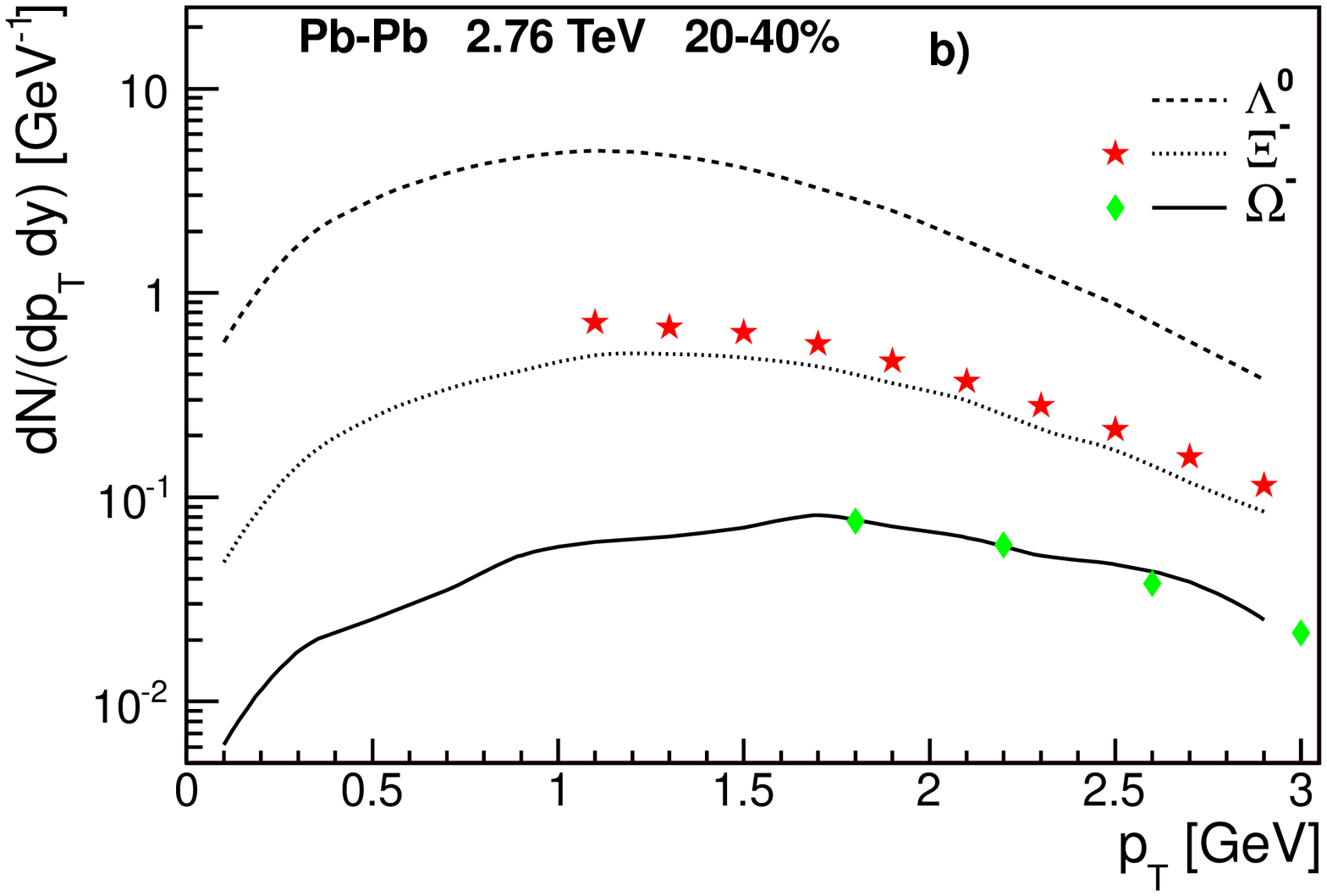}
\caption{(Color online) Transverse momentum spectra of  $\Lambda^0$ (dashed line), $\Xi^-$ 
(dotted line and stars), $\Omega^-$ (solid line and diamonds) in  Pb-Pb collisions with centralities
 $0-20$\% (panel a)  and $20-40$\% (panel b). The lines and the symbols 
represent the results of the  viscous hydrodynamic model and
the  preliminary data of the ALICE Collaboration \cite{Preghenella:2011jv} respectively. }
\label{fig:ex}
\end{figure}

\begin{figure}
\includegraphics[width=.48\textwidth]{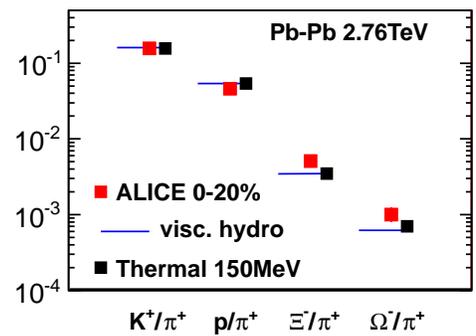}
\caption{(Color online) Ratios of particle yields
 in Pb-Pb collisions with centralities $0-20$\%,
preliminary data of the ALICE Collaboration (squares)   \cite{Preghenella:2011jv} 
are compared to results of  viscous hydrodynamic calculations (lines). }
\label{fig:ratios}
\end{figure} 

The production rate of strange baryons with higher masses $\Xi$ and $\Omega$
is very sensitive to the chemical freeze-out temperature. In Fig. \ref{fig:ex} are shown the
 transverse momentum spectra of $\Lambda^0$, $\Xi^-$ and $\Omega^-$ particles for two centrality classes.
The results of the  viscous hydrodynamic model are in qualitative 
 agreement with the
preliminary data of the ALICE Collaboration. Nonequilibrium corrections in the expanding fireball
increase the effective chemical freeze-out temperature. 
However, the effect is not strong enough to reproduce to observed 
yields of heavy barions. 
It is instructive to look at the ratios of $p_T$ integrated particle yields. The hydrodynamic 
model reproduces  qualitatively  the  observed particles ratios  
(Fig. \ref{fig:ratios}), although the
 nonequilibrium effects are described with only one parameter, the bulk viscosity coefficient.
Deviations from chemical equilibrium at the freeze-out temperature $T_f=140$~MeV 
shift the ratios of heavy particle yields to pion yields up. The ratio $K/\pi$  is very well reproduced, 
the ratio $p/\pi$ is overpredicted by $17\%$ and the ratios
 $\Xi/\pi$, $\Omega/\pi$ are underpredicted by 
$30-40\%$. The ratios obtained from the hydrodynamic simulation with $T_f=140$MeV correspond aproximately to a chemical 
equilibrium temperature of $150$MeV (Fig. \ref{fig:ratios}). The deviations of the calculated particles ratios from the 
observations 
indicate that the data cannot be described using a single chemical freeze-out temperature, as mentioned above, the agreement 
could be improved using different equilibration rates for different particle species.

\begin{figure}
\includegraphics[width=.44\textwidth]{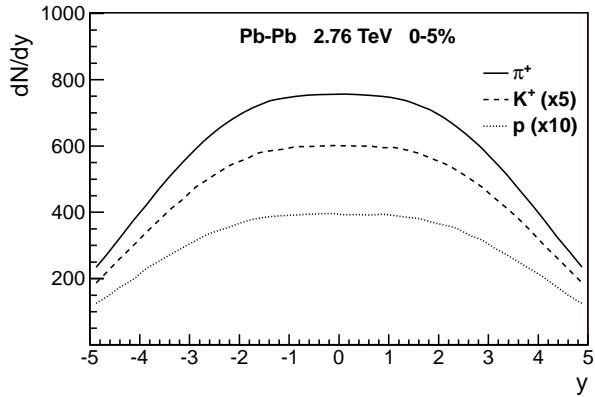}
\caption{Rapidity distributions of $\pi^+$ (solid line), K$^{+}$ (dashed line) and p (dotted line)
 emitted in Pb-Pb collisions with centralities $0-5$\%, calculated in the $(3+1)$-D 
viscous hydrodynamic model.}
\label{fig:dndy}
\end{figure}

The $(3+1)$-D hydrodynamic model does not assume boost-invariance and gives predictions on the rapidity 
dependence of particle spectra. In practice, the initial density in the longitudinal direction is 
adjusted to reproduce the final pseudorapidity distribution (Fig. \ref{fig:dndeta}). Once the 
transverse momentum spectra and the pseudorapidity distributions are found to be in agreement 
with the experimental observations, the rapidity distributions of identified particles can be reliably 
estimated. The rapidity distributions for pions, kaons, and protons are shown in Fig. \ref{fig:dndy}.
We find that the distributions are not boost-invariant. However, in the range $|y|<1$ the dependence 
of the particle densities on 
rapidity  is very weak. This indicates that $(2+1)$-D hydrodynamic models
represent  a good approximation for the dynamics and the mechanism of particle production in Pb-Pb collisions at 
the LHC at midrapidity.

\begin{figure}
\includegraphics[width=.49\textwidth]{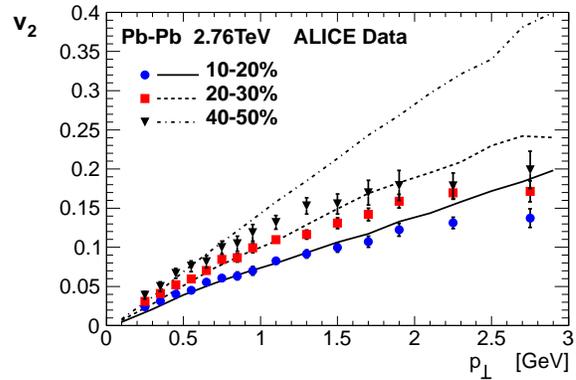}
\caption{ Elliptic flow coefficient of charged particles as function 
of transverse momentum for three centrality classes 
calculated in the viscous hydrodynamic model (lines), the symbols represent the experimental results of the ALICE Collaboration \cite{Aamodt:2010pa}.}
\label{fig:v2ch}
\end{figure}

\begin{figure}
\includegraphics[width=.44\textwidth]{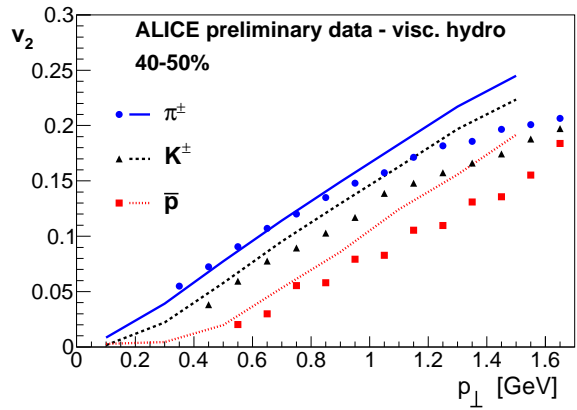}
\caption{Elliptic flow coefficient of identified particles as function of transverse momentum calculated in the viscous hydrodynamic model (lines), the symbols represent the preliminary  results of the ALICE Collaboration \cite{Krzewicki:2011ee}, centrality $40-50$\%.}
\label{fig:v2id}
\end{figure}

Using the hydrodynamic model with optical Glauber initial conditions, 
one can calculate the elliptic flow coefficient.
The elliptic flow coefficient for charged particle at different 
centralities is presented in Fig. \ref{fig:v2ch}. The calculation 
describes well the data for soft momenta. The results  are consistent with predictions of other viscous hydrodynamic  codes 
\cite{Luzum:2010ag,Shen:2011eg,Song:2011qa,Bozek:2011wa,Schenke:2011tv,Qiu:2011hf,Retinskaya:2012ky,Niemi:2012ry}.
Calculations of the elliptic flow for central collisions or of the triangular flow require  event-by-event simulations 
including fluctuations in the initial state, and are outside the scope of this paper.
An observable related to the freeze-out conditions for specific particles is
 the elliptic flow of identified particles.
 As can be seen in Fig. \ref{fig:v2id}, the hydrodynamic model with bulk viscosity corrections and freeze-out at $140$MeV gives a
slightly too small splitting between the elliptic flow of pions, kaons, and protons.

\section{Discussion}

We present  $(3+1)$-D viscous hydrodynamic calculations of the spectra of particles emitted in 
Pb-Pb collisions at $\sqrt{s_{NN}}=2.76$~TeV. We find that the transverse momentum spectra are sensitive 
to  bulk viscosity effects. Bulk viscosity in an exploding fireball leads to deviations from 
local equilibrium in the fluid elements. With the bulk viscosity coefficient $\zeta/s=0.08$ and 
a freeze-out temperature $T_f=140$~MeV we find a satisfactory agreement with preliminary 
experimental data  
for the transverse momentum
 spectra of pions, kaons and protons. At the same time, the size of the fireball and the amount 
of the collective transverse flow of the fluid accumulated in the hydrodynamic phase 
are compatible with the experimental measurements 
of the momentum dependence of the interferometry radii $R_{out}$, $R_{side}$, and $R_{long}$.

Transverse momentum spectra of identified particles in semi-central collisions are reproduced as well.
In particular, we find a strong transverse push visible in the 
$p_T$ spectra of protons. On the other hand, 
bulk viscosity effects make the pion spectra softer, in agreement with the experiment. 
Using the same freeze-out conditions we calculate the $\Xi^-$ and $\Omega^-$ spectra and compare to 
the preliminary data of the ALICE Collaboration for 
 two different centrality classes.
Nonequilibrium 
corrections make the effective chemical freeze-out temperature to be $150$~MeV and not $140$~MeV 
as given by the energy density at freeze-out. The ratios of particle yields  $K/\pi$ are very well reproduced, 
while the ratio $p/\pi$ is slightly overpredicted, and the strange barion abundances are underpredicted. This shows 
that an effective single
chemical freeze-out temperature (generated dynamically in viscous hydrodynamics)
 cannot describe all the measured particle ratios. The 
elliptic flow of identified particles from the hydrodynamic model 
shows a splitting according to the particle mass,  but  slightly 
smaller than observed 
experimentally.
Distributions of identified particles in rapidity show that the system is not boost-invariant. 
However, in the limited interval  of central rapidities $|y|<1$  an approximate plateau is seen
 in the  rapidity distributions.

\section*{Acknowledgment}
The work is supported  by the Supported by Polish Ministry of Science and Higher Education, grants N~N202~263438 and N~N202 086140.

\bibliography{../hydr}

\end{document}